\documentclass[%
 reprint,
 amsmath,amssymb,
 aps,
prl,
floatfix,
]{revtex4-2}

\usepackage{graphicx}
\usepackage{dcolumn}
\usepackage{bm}
\usepackage{amsmath,amssymb}

\usepackage{tikz}

\usepackage{multirow}
\usepackage{array}
\usepackage{cancel}

\usepackage[colorlinks=true,bookmarks=false,citecolor=blue,urlcolor=blue]{hyperref}

\begin{document}

\title{Stochastic Dynamics of Skyrmions on a Racetrack: Impact of Equilibrium and Nonequilibrium Noise}
%\title{Stochastic dynamics (and first passage properties (?)) of skyrmions on the racetrack}

\author{Anton V. Hlushchenko$^1$}
\author{Mykhailo I. Bratchenko$^1$}
\author{Aleksei V. Chechkin$^{1,2,3}$}
 \email{achechkin@mpi-halle.mpg.de}
\affiliation{$^1$National Science Center "Kharkiv Institute of Physics and Technology", 61108 Kharkiv, Ukraine}
\affiliation{$^2$Max Planck Institute of Microstructure Physics, 06120 Halle, Germany}
\affiliation{$^3$Faculty of Pure and Applied Mathematics, Wrocław University of Science and Technology, 50-370 Wrocław, Poland}

%\title{Thermal and spin current noise with SOT current injection in racetracks with domain walls and skyrmions}
\author
\author

\date{\today}

\begin{abstract}
Current-driven motion of domain walls and skyrmions is central to the operation of non-volatile magnetic memory devices. Racetrack memory requires current densities high enough to generate velocities above 50 m/s, but such conditions also enhance spin-current noise. We develop a theoretical framework based on the stochastic Thiele equation to analyze the effects of equilibrium (thermal) and nonequilibrium (spin-current) fluctuations on skyrmion dynamics. From this approach, we derive diffusion coefficients and mean-squared displacements that quantify stochastic motion under both noise sources. Micromagnetic simulations and analytical results demonstrate that spin-current noise dominates skyrmion dynamics in typical racetrack structures up to room temperature. We further address the first-passage-time problem, obtaining the mean first-passage time and its standard deviation along and across the racetrack. These results quantify how random displacements affect skyrmion propagation and detection, providing insights into error sources in high-speed racetrack memory devices.
\end{abstract}

\maketitle
%\section{Introduction}

Developments in the controlled motion of magnetic textures, such as domain walls on racetracks driven by electric currents, have opened the way to alternative non-volatile memory devices with significantly higher performance than conventional solid-state memories \cite{Parkin_2008}. Advances in racetrack memory technology now enable a wide range of applications, from ultra-fast to ultra-dense data storage \cite{Parkin2015, Parkin_2020}.

An alternative to domain-wall magnetic textures are skyrmions \cite{skyrmion_2006}, which can provide denser information storage due to their nanoscale size (down to a few nanometers). These particle-like spin configurations can exist either in magnetic materials with non-centrosymmetric crystal lattices or at the interfaces of magnetic films. Correspondingly, two types of skyrmion states arise from different forms of the Dzyaloshinskii–Moriya interaction (DMI). In non-centrosymmetric lattices \cite{bulk_2009,bulk_2012,Nagaosa2013}, the bulk DMI stabilizes vortex-like (Bloch-type) skyrmions \cite{Sun_2013}. In contrast, interfacial DMI leads to the stabilization of “hedgehog” (Néel-type) skyrmions \cite{interfacial_2008,interfacial_2011}. Moreover, Néel-type skyrmions can also be stabilized in centrosymmetric ferromagnets by imposing a vertical strain gradient via epitaxial growth \cite{Parkin_skyrmions}.

Historically, the first racetrack memories were realized in permalloy nanowires \cite{Parkin_2008}, where domain walls were driven by nanosecond current pulses via the spin-transfer torque (STT) mechanism \cite{Slonczewski_1996,Zhang_2004_PhysRevLett,Thiaville_EuroLett_2005,Barnas_2006_PhysRevB}.
The next generation of racetracks \cite{Sampaio2013,Parkin2015,Chubykalo_Nat_2020} was achieved by injecting a spin current from a heavy-metal layer into an adjacent ferromagnetic layer. The resulting spin-current–induced motion of the magnetic texture arises from spin–orbit scattering of conduction electrons via the spin Hall effect in the heavy-metal layer \cite{Hirsch_1999,Miron2010,Qiming_2021} and is described by the spin–orbit torque (SOT) mechanism. In this work, we focus on racetracks with SOT current injection, which enables higher velocities of magnetic textures. However, our theoretical framework can be readily extended to the STT case.

Real magnetic systems are inherently subject to random fluctuations, which may include both equilibrium and nonequilibrium sources of noise. Equilibrium thermal agitation leads to fluctuations of the magnetization field \cite{Brown_PR_1963,Kubo_spin_1970,Garanin_PRB_1997,Garcia_diffusion_1998,Chubykalo_PhysRevB_2012,Chotorlishvili_2013_PhysRevB,Chotorlishvili_2014_PhysRevB,Vedmedenko2019,Nowak_2021}. In contrast, the presence of an external current gives rise to nonequilibrium noise \cite{Mishchenko_PRB_2003,Belzig_PRB_2004,Sauret_PRL_2004,Wang_PRB_2004,Foros_2005,Dragomirova_PRB_2007,Chudnovskiy_2008,Meair_PRB_2011,Chudnovskiy_shot_2014,Barnas_2016_PhysRevB,Bratas_2018}.
In high-speed spintronic devices (with typical velocities of at least about $50$ m/s or higher \cite{Parkin2015,woo2016observation,Vedmedenko_2020,da2025neuromorphic}), large current densities can lead to enhanced spin-current fluctuations. In conventional electrical conductors, such as heavy-metal layers interfaced with nanoscale ferromagnets, spin-current noise increases magnetization noise through fluctuating spin torque \cite{Foros_2005,Bratas_2018}. Anomalous behavior of magnetization fluctuations in current-controlled spin valves has already been demonstrated at cryogenic temperatures \cite{Zholud_PRL_2017}, and this effect is expected to remain significant even at room temperature.

One of the major technological challenges in racetrack memory is achieving the highest possible packing density of nanoscale magnetic textures \cite{JaeChun_ACSNANO_2024,JaeChun_Science_2024,Ishibashi_SciAdv_2024}, while still enabling reliable detection of domain walls or skyrmions when they are densely packed. In this work, we focus on describing the effects of both thermal noise and spin-current noise on skyrmion dynamics. Our results reveal a pronounced impact of spin-current fluctuations on the stable operation of racetrack memory devices. For typical racetrack geometries and operational conditions, trajectory offsets caused by spin-current noise dominate over thermal noise.
This stochastic spreading of skyrmion trajectories is quantified in terms of the mean squared displacement (MSD), which we obtain analytically using the Thiele approach and numerically through micromagnetic simulations. We also address the first-passage-time (FPT) problem for skyrmion motion on the racetrack, estimating the mean and standard deviation of the random times at which a skyrmion reaches either the racetrack side boundary or the detection region. Interaction of a skyrmion with a boundary can lead to its annihilation \cite{Jiang2017} and consequent loss of stored information, thereby compromising racetrack stability. Conversely, time uncertainty in reaching the detection region affects signal decoding quality.
A deeper understanding of the roles of equilibrium and nonequilibrium fluctuations in skyrmion motion is therefore essential for the design of next-generation, high-speed, and high-density spintronic devices.

%\section{Theory}

The dynamics of magnetization  $\textbf{M}(\textbf{r},t)=(M_1(\textbf{r},t),M_2(\textbf{r},t),M_3(\textbf{r},t))$ is described by the Landau-Lifshitz-Gilbert (LLG) equation \cite{Landau_PhysSowjet_1935,Gilbert_IEEE_2004}, which includes additional term accounting for the spin transfer moment:
\begin{equation}
\begin{split}\label{LLG_torque}
\partial_t\textbf{M}=
-\gamma\textbf{M}\times\textbf{H}_\text{eff}+\frac{\alpha}{M_s}\textbf{M}\times\partial_t\textbf{M}+\textbf{T}_\text{SOT}.
\end{split}
\end{equation}
Here $\textbf{H}_\text{eff}$ is the effective field, $\gamma$ is the gyromagnetic ratio, $\alpha$ is the damping parameter, $M_s$ is
the saturation magnetization. 
The spin-orbit torque $\textbf{T}_\text{SOT}$ can be expressed as a combination of  Slonczewski’s torque \cite{Slonczewski_1996} and the damping-like torque:
\begin{equation}
\begin{split}\label{Slonczewski}
\textbf{T}_\text{SOT}=-\frac{\gamma}{M_s^2\mathcal{V}}\textbf{M}\times\left(\textbf{M}\times\textbf{I}_s\right)-\frac{\gamma\beta}{M_s\mathcal{V}}\textbf{M}\times\textbf{I}_s,
\end{split}
\end{equation}
where $\textbf{I}_s=\textbf{P}I\hbar/(2e)$ $(\textbf{P}=P\cdot\textbf{p}=P\cdot(\cos{\psi},\sin{\psi},0))$ the spin-polarization vector, $\hbar$ the reduced Planck constant, $e$ the electron charge, $I$ the electric current, $\beta$ the amplitude of the damping-like torque, and $\mathcal{V}$ the sample volume.  

Next, we generalize the LLG equation to account for both thermal noise and spin-current noise.  Following Ref.~\cite{Brown_PR_1963}, the thermal noise is represented by a fluctuating magnetic field $\delta\textbf{h}(\textbf{r},t)$ with zero mean as $\textbf{H}_\text{eff}\rightarrow\textbf{H}_\text{eff}+\delta\textbf{h}$. In addition to the thermal noise, by analogy with Ref.~\cite{Chudnovskiy_2008}, we introduce spin-current noise $\delta\textbf{I}(\textbf{r},t)$ as $\textbf{I}_s\rightarrow\textbf{I}_s+\delta\textbf{I}$. This form of noise is analogous to the charge shot noise \cite{BLANTER20001}. The discrete nature of the electron transport in the heavy metal leads to stochastic angular-momentum transfer between conduction electrons and atomic spins in the ferromagnetic layer \cite{Foros_2005,Chudnovskiy_shot_2014}. 
The stochastic LLG equation reads as 
\begin{equation}
\begin{split}\label{LLG_torque_current_thermal}
\partial_t\textbf{M}=
-\gamma\textbf{M}\times(\textbf{H}_\text{eff}+\delta\textbf{h})+\frac{\alpha}{M_s}\textbf{M}\times\partial_t\textbf{M}\\
-\frac{\gamma}{M_s^2\mathcal{V}}\textbf{M}\times\left(\textbf{M}\times(\textbf{I}_s+\delta\textbf{I})\right)-\frac{\gamma\beta}{M_s\mathcal{V}}\textbf{M}\times(\textbf{I}_s+\delta\textbf{I}).
\end{split}
\end{equation}
The autocorrelation functions of fluctuation terms take the form \cite{Brown_PR_1963,Foros_2005,Bratas_2018}
\begin{equation}
\begin{split}\label{current_thermal_noise_}
\langle{\delta{h}}_{i}(t,\textbf{r}){\delta{h}}_{j}(t',\textbf{r}')\rangle=\frac{2{\alpha}k_BT}{\gamma{M}_s}\delta_{ij}\delta(\textbf{r}-\textbf{r}')\delta(t-t'), \\ \langle{\delta{I}}_{i}(t,\textbf{r}){\delta{I}}_{j}(t',\textbf{r}')\rangle=\frac{\alpha{M}_s\mathcal{V}^2}{\gamma}\delta_{ij}\delta(\textbf{r}-\textbf{r}')\delta(t-t') \\
\times\frac{\delta\varepsilon}{E_F}(eU\coth{\frac{eU}{2k_BT}}-2k_BT). 
\end{split}
\end{equation}
Here $i,j=\{x,y,z\}$, $\delta_{ij}$ is the Kronecker delta function, $\delta(\cdot)$ is the Dirac delta function,  $U$ and $T$ denote the applied voltage and temperature, $E_F$ is the Fermi energy in the heavy-metal layer, $\delta\varepsilon$ is the exchange splitting in the ferromagnetic layer, which can be estimated as $\delta\varepsilon\sim{E_F}/5$ \cite{Foros_2005}. We consider equilibrium and nonequilibrium noise sources of different origins, assuming that they are uncorrelated, $\langle{\delta{h}}_{i}(t,\textbf{r}){\delta{I}}_{j}(t',\textbf{r}')\rangle=0$. 
In the zero temperature limit, the current fluctuations persist, while in the absence of an applied voltage, only thermal fluctuations remain.
\begin{figure*}
\centering
\includegraphics[width=\linewidth]{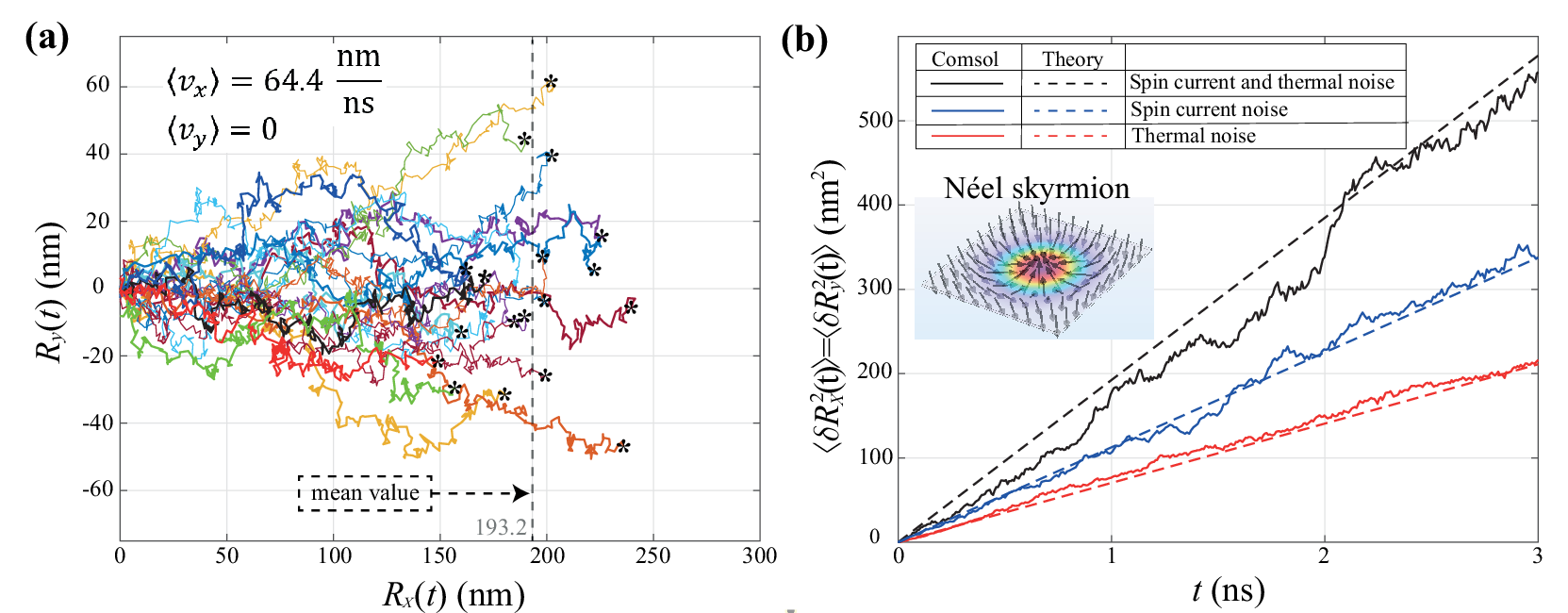}
\caption{\label{Displacements} (a) Example trajectories of Néel skyrmion motion under the combined influence of thermal and spin-current noise. The racetrack width of $150$ nm is chosen to be sufficiently large to prevent skyrmion interaction with the racetrack boundaries during a $3$ ns current pulse.
 (b) MSD of the skyrmion position obtained from both micromagnetic simulations (solid lines, $100$ trajectories) and the analytical calculations (dashed lines). The results are shown for the cases with only thermal noise, only spin-current noise, and both noise sources (from bottom to top, respectively).}
\end{figure*}

We employ the Thiele approach to evaluate the contribution of equilibrium and nonequilibrium noise to skyrmion displacements. The Thiele formalism \cite{Thiele_PRL_1973} uses the traveling-wave Ansatz $\textbf{M}(\textbf{r},t)=\textbf{M}(\textbf{r}-\textbf{R}(t))$, where $\textbf{R}(t)$ is the position of the center of skyrmion.  In our study, we neglect the higher derivatives of $\textbf{R}(t)$ \cite{Kamppeter_PRB_1999}, which have only a minor effect on the main results, and reduce Eq.~(\ref{LLG_torque_current_thermal}) to the form of the stochastic Thiele equation (see details in the Supplemental Material \cite{Sup_mat}, Section I)  
\begin{equation}\label{Thiel_torque_shot_2}
\begin{split}
&\widehat{G}\cdot\dot{\textbf{R}}+\alpha\widehat{D}\cdot\dot{\textbf{R}}+\textbf{F}=0, \\
&\textbf{F}=\textbf{F}_{eff}+\textbf{F}_\text{SOT}+\textbf{F}_{\delta{h}}+\textbf{F}_{\delta{I}},
\end{split}
\end{equation}
where $\widehat{G}$ is the gyrotropic tensor and $\widehat{D}$ is the dissipation tensor, which are given by $G_{ij}=-{1}/{(\gamma{M}_s^2)}\int{\textbf{M}\cdot(\nabla_i\textbf{M}\times\nabla_j\textbf{M})d\textbf{r}}$ and ${D}_{ij}={1}/{(\gamma{M}_s)}\int{\nabla_i\textbf{M}\cdot\nabla_j\textbf{M}d\textbf{r}}$, respectively. The effective force is $F_{eff,i}=\int{\textbf{H}_{eff}\cdot\nabla_i{\textbf{M}}d\textbf{r}}$, and $F_{\text{SOT},i}={1}/({\gamma{M}_s^2})\int{(\textbf{M}\times\nabla_i\textbf{M})\cdot{\textbf{I}}_sd\textbf{r}}+{\beta}/({\gamma{M}_s})\int{{\textbf{I}_s}\cdot\nabla_i{\textbf{M}}d\textbf{r}}$. The two uncorrelated stochastic forces take the form
\begin{equation}\label{stochastic_forces}
\begin{split}
&F_{\delta{I},i}=\frac{1}{\gamma{M}_s^2}\int{(\textbf{M}\times\nabla_i\textbf{M}+M_s\beta\nabla_i{\textbf{M}})\cdot\delta{\textbf{I}}d\textbf{r}},\\
&F_{\delta{h},i}=\int{\delta\textbf{h}\cdot\nabla_i{\textbf{M}}d\textbf{r}},
\end{split}
\end{equation}
and their autocorrelation functions are given by
\begin{equation}\label{cor_2}
\begin{split}
&\langle{F}_{\delta{h},i}(t)F_{\delta{h},j}(t)\rangle=\alpha{k_B}T{D}_{ij}\delta(t-t'), \\
&\langle{F}_{\delta{I},i}(t)F_{\delta{I},j}(t)\rangle=\alpha(1+\beta^2){D}_{ij}\delta(t-t') \\
&\times\frac{\delta\varepsilon}{E_F}(eU\coth{\frac{eU}{2k_BT}}-2k_BT).
    \end{split}
\end{equation}
The resulting Eqs.~(\ref{Thiel_torque_shot_2}-\ref{cor_2}) describe two-dimensional skyrmion diffusion in thin ferromagnetic films in the ($x$-$y$) plane. Owing to the rotational symmetry of a skyrmion at rest  we neglect ${D}_{xy}$ and ${D}_{yx}$, and set ${D}_{xx}={D}_{yy}=D$ and ${G}_{xy}=G$ \cite{Thiaville_2018}, which is also confirmed by our micromagnetic simulations. The solution of Eq.~(\ref{Thiel_torque_shot_2}) for the velocity $\dot{{\textbf{R}}}(t)=\textbf{v}(t)=\{v_x(t),v_y(t)\}$ of the skyrmion is given by
\begin{equation}\label{velocity_skyrmion}
\begin{split}
&v_x=\frac{-GF_y - \alpha{D}F_x}{G^2 + \alpha^2D^2},\quad v_y=\frac{GF_x-\alpha{D}F_y}{G^2 + \alpha^2D^2}.
\end{split}
\end{equation}
After evaluating the velocity correlation functions and integrating over time $\textbf{R}(t)=\int_0^t{\textbf{v}(t')dt'}$ (see details in the Supplemental Material \cite{Sup_mat}, Section II) one obtains the MSD $\langle\delta\textbf{R}^2(t)\rangle=\langle(\textbf{R}(t)-\langle\textbf{R}(t)\rangle)^2\rangle$,
\begin{equation}\label{skyrmion_diffusion}
\begin{split}
&\langle\delta{R}^2_x(t)\rangle=\langle\delta{R}^2_y(t)\rangle=2\mathcal{D}t,
\end{split}
\end{equation}
where the total diffusion coefficient $\mathcal{D}=\mathcal{D}_T+\mathcal{D}_I$ accounts for the respective contributions of thermal and spin current noise fluctuations,
 \begin{equation}\label{skyrmion_diffusion_1}
\begin{split}
&\mathcal{D}_T=\frac{\alpha{D}}{2(G^2+\alpha^2D^2)}2k_BT,\\
&\mathcal{D}_I=\frac{\alpha{D}(1+\beta^2)}{2(G^2+\alpha^2D^2)}\frac{\delta\varepsilon}{E_F}(eU\coth{\frac{eU}{2k_BT}}-2k_BT).
\end{split}
\end{equation}
Note the non-monotonic dependence of the diffusion coefficient $\mathcal{D}$ on the damping parameter $\alpha$, which results in diffusion suppression as $\alpha\rightarrow 0$. This behavior is characteristic of skyrmions, in contrast to domain walls, for which $G$= 0, and  $\mathcal{D}\sim1/\alpha$ \cite{Schutte_PRB_2014,Thiaville_2018}. 

Micromagnetic simulations were performed using the micromagnetic module of  COMSOL Multiphysics \cite{comsolMicromag,micromagModule} for a two-dimensional system. A comparison between this module and mumax$^3$ \cite{mumax3}  is presented in the Supplemental Material \cite{Sup_mat}, Section~III.  We extended the COMSOL module with a user-defined code that incorporates thermal and spin-current fluctuations in accordance with Eqs.~(\ref{current_thermal_noise_}).
The following parameters of the effective field $\textbf{H}_\text{eff}$ (see Supplemental Material, Section~III) were used in the simulations: exchange coefficient $A_\text{ex}=10^{-11}$~J/m, anisotropy $K_u=10^6$~J/m$^3$, saturation magnetization $M_s=0.6\cdot10^{6}$~A/m, interfacial DMI constant $D_\text{DMI}=-3.5\cdot10^{-3}$~J/m$^2$, and damping parameters $\alpha=\beta=0.3$. The racetrack sample has an ($x$-$y$) cross section of $500\times150$~nm$^2$ and thickness $\Delta_f=1$ nm. The applied current density is $J=4\cdot10^{11}$~A/m$^2$, the voltage is $U=0.5$~V, and a linear current-voltage relationship $ {J}\propto{U}$ is assumed. The temperature is set to $T=300$~K.  These parameters are typical of experimental racetrack systems \cite{Sampaio2013,JaeChun_ACSNANO_2024,JaeChun_Science_2024}. The topological properties of skyrmions are fully characterized by three quantities ($Q, Q_v, Q_h$), representing the topological charge, vorticity number, and helicity number, respectively \cite{Tretiakov_PhysRevB_2017,Zhang_2020}. We restrict our analysis to the DMI, that stabilizes N\'eel-type skyrmions with ($1,1,\pi$). 

\begin{figure*}
\centering
\includegraphics[width=\linewidth]{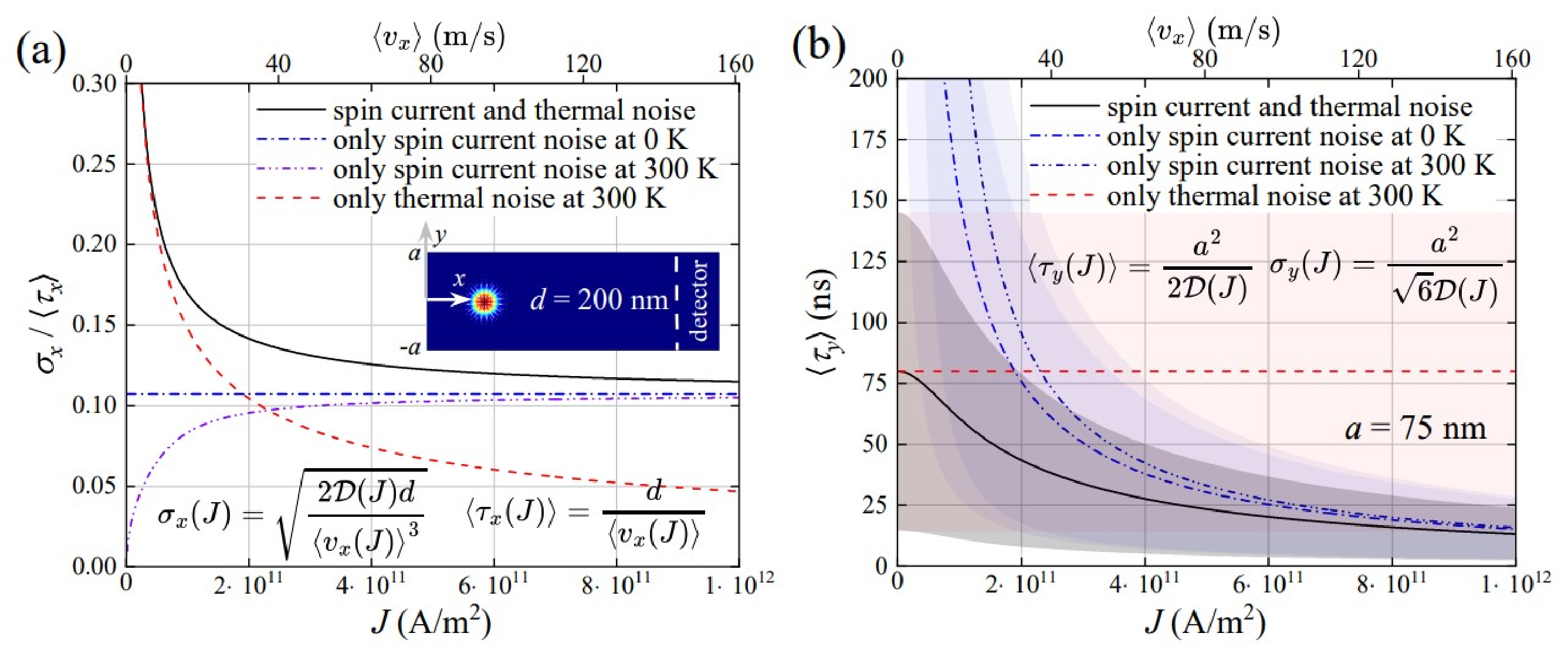}
\caption{\label{FPT} (a) Normalized SD of the FPT  to reach the detector position ($x$-direction), and (b) Mean FPT  to reach the racetrack boundary $\pm{a}$ ($y$-direction) as functions of the current density $J$ for various combinations of noises.  Shaded regions in panel (b) indicate the standard deviations of the mean FPT along the $y$-axis. Here $d$ = 200 nm, $a$ = 75 cm.}
\end{figure*}

Using micromagnetic modeling, the gyrocoupling constant $G$ and dissipation tensor component $D$ take the following values: $G=4\pi{M}_s\Delta_f/\gamma$ and $D=18.1M_s\Delta_f/\gamma$ (see details in the Supplemental Material, Section III). In this work, we focus on the current-driven motion of skyrmions  along the $x$-axis, which can be achieved by suppressing the transverse motion along the $y$-axis induced by the skyrmion Hall effect \cite{Jiang_Nature_2017}.
Following Ref.~\cite{Gobel_PRB_2019}, the condition $\langle{v}_y\rangle=0$  can be satisfied by choosing the angle $\psi$  of the polarization vector $\textbf{p}$ as
\begin{equation}\label{suppress_Hall_effect}
\begin{split}
\psi=\arctan{\frac{G}{\alpha{D}}}+\frac{\pi}{2}-Q_h.
\end{split}
\end{equation}
For the N\'eel skyrmion, this yields $\psi=-23.36^{\circ}$, which is consistent with our micromagnetic simulations (note that the angle $\delta$ in Ref.~\cite{Gobel_PRB_2019} corresponds to $\psi$ as $\delta=\psi-\pi/2$). For the chosen set of parameters the resulting drift velocity along the $x$-axis is estimated as $\langle{v}_x\rangle=64.4$ m/s (see Supplemental Material, Section III).

Exemplary skyrmion trajectories obtained from micromagnetic simulations using COMSOL multiphysics are shown in Fig.~\ref{Displacements}(a), where both thermal and spin-current noises are taken into account. The current pulse duration is set to 3 ns, resulting in a mean displacement along the $x$-axis of 193.2 nm. The final skyrmion positions after 3 ns are marked by black asterisks. It can be seen that thermal and spin-current fluctuations cause significant random deviations around the mean position in both the $x$- and $y$-directions.   
The MSDs along both axes, obtained from simulated trajectories, are shown by solid lines in Fig.~\ref{Displacements}(b) together with the corresponding MSDs calculated from the Thiele approach, Eqs.~(\ref{skyrmion_diffusion}, \ref{skyrmion_diffusion_1}), shown by the dashed lines. The Figure demonstrates excellent agreement between the micromagnetic simulations and the analytical predictions based on the Thiele formalism. For the selected system parameters, nonequilibrium fluctuations dominate over equilibrium (thermal) fluctuations.

We now proceed to study the first passage problem for skyrmions on the racetrack in the situation where the skyrmion Hall effect and therefore the transverse drift across the racetrack, are absent.
Using Eqs.~(\ref{Thiel_torque_shot_2}-\ref{skyrmion_diffusion_1}) we rewrite the Thiele equation (\ref{Thiel_torque_shot_2}) in the form of a standard stochastic equation of motion
\begin{equation}\label{difura}
\begin{split}
&\dot{\textbf{R}}=\langle\textbf{v}\rangle+\sqrt{2\mathcal{D}}\bm{\xi}(t).
\end{split}
\end{equation}
Here,  $\langle\textbf{v}\rangle=\{\langle v_x \rangle$, 0\} is the mean drift velocity, whose components are determined from Eq.~(\ref{velocity_skyrmion}), and $\bm{\xi}(t)$ is a white Gaussian noise with the zero mean and autocorrelation function $\langle\xi_i(t)\xi_j(t)\rangle=\delta_{ij}\delta(t-t')$.
The form (\ref{difura}) is particularly convenient for analyzing first passage phenomena \cite{CoxMiller_2001,Redner_2001}. Here, we focus on the mean FPTs $\langle\tau_{x,y}\rangle$ along and across the racetrack, as well as their standard deviations (SDs) $\sigma_{x,y} = \sqrt{\langle\tau^2_{x,y}\rangle - \langle\tau_{x,y}\rangle^2}$. The details of derivation for semi-infinite and finite domains are provided in the Supplemental Material, Section IV. Figure ~\ref{FPT}(a) illustrates the nontrivial behavior of the normalized SD $\sigma_x /\langle\tau_x\rangle =\sqrt{2\mathcal{D}(J)/(\langle{v}_x(J)\rangle d)}$ for the FPT along the racetrack 
 as the function of the current density $J$, where $d$ denotes the distance between the initial position at $t = 0$ and the detector.
 For skyrmion diffusion across the racetrack, the mean FPT to reach  the racetrack side boundaries $\pm{a}$ and its SD are given by $\langle\tau_y\rangle=a^2/(2\mathcal{D})$ and $\sigma_y=a^2/(\sqrt{6}\mathcal{D})$, respectively, differing only by a constant prefactor.
Figure ~\ref{FPT}(b) presents the mean FPT $\langle\tau_y(J)\rangle$ across the racetrack with ${a}=75$ nm as a function of the current density $J$ in the presence of thermal and/or spin current fluctuations. 

The analytical results presented in Fig.~\ref{FPT}(a) indicate that the spread of the mean FPT along the racetrack is governed by thermal fluctuations at low drift velocities, whereas at higher, yet still experimentally typical, velocities for present-day racetracks, non-equilibrium spin-current fluctuations become dominant. Remarkably, at zero temperature the normalized standard deviation of the FPT is independent of the current density. As the current density increases, the normalized total spread of the mean FPT approaches a constant value determined by spin-current fluctuations at zero temperature.

Figure~\ref{FPT}(b) shows the mean FPT across the racetrack as a function of the current density. Owing to the suppression of the skyrmion Hall effect, the motion along the $y$-axis is purely diffusive in nature.  Similar to the motion along the racetrack, at low current densities (corresponding to low drift velocities) the mean FPT in the transverse direction is governed by thermal fluctuations, whereas at higher currents, spin-current fluctuations dominate the process of reaching the boundaries. For the parameters used here, the mean FPT along the $y$-axis exceeds the typical duration of the current-pulse sequence \cite{JaeChun_Science_2024}; however, hitting the boundary can become significant for narrow racetracks due to the quadratic dependence on the racetrack width. 
%

%\section{Conclusions}

We developed an analytical framework to describe the combined effects of equilibrium (thermal) and nonequilibrium (spin-current) fluctuations on skyrmion dynamics in racetrack systems. Using the Thiele approach, we reduced the stochastic Landau–Lifshitz–Gilbert equation to a Langevin-like form for the skyrmion center of mass, allowing analytical evaluation of the mean squared displacement and first-passage time. The skyrmion exhibits normal diffusion with a coefficient governed by thermal fluctuations at low current densities, while spin-current noise dominates at higher, yet experimentally relevant, currents. 
Even in the absence of the skyrmion Hall effect, diffusive motion can drive the skyrmion to the racetrack boundaries, with the mean FPT across the racetrack becoming comparable to the measurement time for low currents and narrow geometries. 
Micromagnetic simulations including both fluctuation sources confirm these analytical results.

Random displacements of magnetic textures represent a key source of bit-position and timing errors in high-speed racetrack memory devices \cite{Sun_IEEE_2013, Zhang_2015, Chee_IEEE_2018, Ishibashi_SciAdv_2024}. Beyond homogeneous systems, spatial disorder in the energy landscape \cite{Jiang2010, Lee_2011} arising  from variations in magnetic anisotropy or DMI can induce pinning and anomalous diffusion \cite{Ravelosona_2007, Barkhausen_PhysRevLett_2019, Kawaguchi2021}. Together with stochastic texture-creation errors \cite{Ishibashi_SciAdv_2024}, these effects call for extending the present theory to include diverse noise mechanisms that ultimately limit the reliability of racetrack operation.

\label{sec:Concl}

%-----------------------------------------------------%
%\section*{Acknowledgments}
%\label{sec:ack}
\emph{Acknowledgments}. The authors acknowledge illuminating discussions with S.S.P. Parkin and J.-C. Jeon, and financial support from the BMBF project 01DK24006 PLASMA-SPIN-ENERGY. AVH also acknowledges support of the STCU grant No. 9918 under the “2025 IEEE  in Ukraine initiative".

\emph{Data availability}. The data that support the findings of this article are openly available \cite{hlushchenko_2025_17658842}.

\bibliography{Stochastic}
%\section*{Appendix A: Slonczewski’s model}\label{Slonczewski_mod}

%

\onecolumngrid
\section*{Supplementary material}
This Supplemental Material provides detailed derivation of the Thiele equation, Section I, and fluctuation forces and diffusion
coefficients of the skyrmion, Section II; details of the micromagnetic simulations, Section III; and derivation of the first passage time moments for the  skyrmion motion along and across the racetrack, Section IV.

\section{I. The Thiele Equation}\label{Sup_1}
The stochastic LLG equation is
\begin{equation}
\begin{split}\label{LLG_torque_current_thermal}
\partial_t\textbf{M}
+\gamma\textbf{M}\times(\textbf{H}_\text{eff}+\delta\textbf{h})-\frac{\alpha}{M_s}\textbf{M}\times\partial_t\textbf{M}+\frac{\gamma}{M_s^2\mathcal{V}}\textbf{M}\times\left(\textbf{M}\times(\textbf{I}_s+\delta\textbf{I})\right)+\frac{\gamma\beta}{M_s\mathcal{V}}\textbf{M}\times(\textbf{I}_s+\delta\textbf{I})=0.
\end{split}
\end{equation}
To study dynamics of magnetic textures, we use the Thiele approach, which employs the traveling-wave Ansatz $\textbf{M}=\textbf{M}(\textbf{r}-\textbf{R}(t))$ and results in $\partial_t\textbf{M}(\textbf{r}-\textbf{R}(t))=-\nabla_i\textbf{M}~\dot{{R}_i}(t)$. In our study, we omit the higher derivatives of $\textbf{R}(t)$ \cite{Kamppeter_PRB_1999}, which affect a small initial time interval (characteristic time amounts to several tens of picoseconds \cite{Thiaville_2018}).
To obtain the Thiele stochastic equation, we begin with Eq.~\ref{LLG_torque_current_thermal}, multiply it by $\frac{1}{M_s^2\gamma}\textbf{M}\cdot[\nabla_i\textbf{M}\times\ldots]$, and then integrate over $\textbf{r}$. We also use the relations $M_i\cdot M_i=M_s^2$, $\nabla_j M_i\cdot M_i=0$, and $\partial_tM_i\cdot M_i=0$ \cite{Thiele_PRL_1973}. Each term in Eq.~\ref{LLG_torque_current_thermal} generates a separate force:

\begin{itemize}
\item The gyrotropic force 
\begin{equation}
\begin{split}\label{gyrotropic_force}
\frac{1}{M_s^2\gamma}\int{\textbf{M}\cdot[\nabla_i\textbf{M}\times}\partial_t\textbf{M}]d\textbf{r}=\frac{-1}{M_s^2\gamma}\int{\textbf{M}\cdot[\nabla_i\textbf{M}\times}\nabla_j\textbf{M}]d\textbf{r}\cdot\dot{{R}_j}=\hat{G}\cdot\dot{\textbf{R}}, G_{ij}=\frac{-1}{M_s^2\gamma}\int{\textbf{M}\cdot[\nabla_i\textbf{M}\times}\nabla_j\textbf{M}]d\textbf{r}.
\end{split}
\end{equation}
  \item The force generated by the effective field 
\begin{equation}
\begin{split}\label{effective_force}
\frac{1}{M_s^2}\int{\textbf{M}\cdot[\nabla_i\textbf{M}\times[\textbf{M}\times\textbf{H}_\text{eff}]]d\textbf{r}}=\frac{1}{M_s^2}\int{\textbf{M}\cdot[\textbf{M}(\nabla_i\textbf{M}\cdot\textbf{H}_\text{eff})-\textbf{H}_\text{eff}(\bcancel{\nabla_i\textbf{M}\cdot\textbf{M}})]d\textbf{r}}=\int{\nabla_i\textbf{M}\cdot\textbf{H}_\text{eff}d\textbf{r}}.
\end{split}
\end{equation}
  
  \item  The thermal fluctuation force
\begin{equation}
\begin{split}\label{thermal_force}
\frac{1}{M_s^2}\int{\textbf{M}\cdot[\nabla_i\textbf{M}\times[\textbf{M}\times\delta\textbf{h}]]d\textbf{r}}=\frac{1}{M_s^2}\int{\textbf{M}\cdot[\textbf{M}(\nabla_i\textbf{M}\cdot\delta\textbf{h})-\delta\textbf{h}(\bcancel{\nabla_i\textbf{M}\cdot\textbf{M}})]d\textbf{r}}=\int{\nabla_i\textbf{M}\cdot\delta\textbf{h}d\textbf{r}}.
\end{split}
\end{equation}
 \item  The dissipation force
 \begin{equation}
\begin{split}\label{dissipation_force}
\frac{-\alpha}{M_s^3\gamma}\int{\textbf{M}\cdot[\nabla_i\textbf{M}\times[\textbf{M}\times\partial_t\textbf{M}]]d\textbf{r}}=\frac{-\alpha}{M_s^3\gamma}\int{\textbf{M}\cdot[\textbf{M}(\nabla_i\textbf{M}\cdot\partial_t\textbf{M})-\partial_t\textbf{M}(\bcancel{\nabla_i\textbf{M}\cdot\textbf{M}})]d\textbf{r}}\\ =\frac{-\alpha}{M_s^3\gamma}\int{M_s^2(\nabla_i\textbf{M}\cdot\partial_t\textbf{M})d\textbf{r}}=\frac{\alpha}{M_s\gamma}\int{\nabla_i\textbf{M}\cdot\nabla_j\textbf{M}d\textbf{r}}\dot{{R}_j}=\alpha\hat{D}\cdot\dot{\textbf{R}}, D_{ij}=\frac{1}{M_s\gamma}\int{\nabla_i\textbf{M}\cdot\nabla_j\textbf{M}d\textbf{r}}.
\end{split}
\end{equation}
 \item  Force induced by Slonczewski's torque term
 \begin{equation}
\begin{split}\label{Slonczewski_force}
\frac{1}{M_s^4\mathcal{V}}\int{\textbf{M}\cdot(\nabla_i\textbf{M}\times[\textbf{M}\times[\textbf{M}\times\textbf{I}_s]])d\textbf{r}}=\frac{1}{M_s^4\mathcal{V}}\int{\textbf{M}\cdot[\textbf{M}(\nabla_i\textbf{M}\cdot[\textbf{M}\times\textbf{I}_s])-[\textbf{M}\times\textbf{I}_s](\bcancel{\nabla_i\textbf{M}\cdot\textbf{M}})]d\textbf{r}} \\=\frac{1}{M_s^2\mathcal{V}}\int{\nabla_i\textbf{M}\cdot[\textbf{M}\times\textbf{I}_s]d\textbf{r}}=\frac{1}{M_s^2\mathcal{V}}\int{[\nabla_i\textbf{M}\times\textbf{M}]\cdot\textbf{I}_sd\textbf{r}}
.
\end{split}
\end{equation}
\item The damping-like torque force
\begin{equation}
\begin{split}\label{damping_torque_force}
\frac{\beta}{M_s^3\mathcal{V}}\int{\textbf{M}\cdot[\nabla_i\textbf{M}\times[\textbf{M}\times\textbf{I}_s]]d\textbf{r}}=\frac{\beta}{M_s^3\mathcal{V}}\int{\textbf{M}\cdot[\textbf{M}(\nabla_i\textbf{M}\cdot\textbf{I}_s)-\textbf{I}_s(\bcancel{\nabla_i\textbf{M}\cdot\textbf{M}})]d\textbf{r}}=\frac{\beta}{M_s\mathcal{V}}\int{\nabla_i\textbf{M}\cdot\textbf{I}_sd\textbf{r}}.
\end{split}
\end{equation}

  \item  The spin current fluctuation force
 \begin{equation}
\begin{split}\label{spin_current_force}
\frac{1}{M_s^4\mathcal{V}}\int{\textbf{M}\cdot(\nabla_i\textbf{M}\times[\textbf{M}\times[\textbf{M}\times\delta\textbf{I}]])d\textbf{r}}+\frac{\beta}{M_s^3\mathcal{V}}\int{\textbf{M}\cdot[\nabla_i\textbf{M}\times[\textbf{M}\times\delta\textbf{I}]]d\textbf{r}} \\=\frac{1}{M_s^4\mathcal{V}}\int{\textbf{M}\cdot[\textbf{M}(\nabla_i\textbf{M}\cdot[\textbf{M}\times\delta\textbf{I}])-[\textbf{M}\times\delta\textbf{I}](\bcancel{\nabla_i\textbf{M}\cdot\textbf{M}})]d\textbf{r}}+\frac{\beta}{M_s^3\mathcal{V}}\int{\textbf{M}\cdot[\textbf{M}(\nabla_i\textbf{M}\cdot\delta\textbf{I})-\delta\textbf{I}(\bcancel{\nabla_i\textbf{M}\cdot\textbf{M}})]d\textbf{r}}\\=\frac{1}{M_s^2\mathcal{V}}\int{\nabla_i\textbf{M}\cdot[\textbf{M}\times\delta\textbf{I}]d\textbf{r}}+\frac{\beta}{M_s\mathcal{V}}\int{\nabla_i\textbf{M}\cdot\delta\textbf{I}d\textbf{r}}=\frac{1}{M_s^2\mathcal{V}}\int{[\nabla_i\textbf{M}\times\textbf{M}]\cdot\delta\textbf{I}d\textbf{r}}
+\frac{\beta}{M_s\mathcal{V}}\int{\nabla_i\textbf{M}\cdot\delta\textbf{I}d\textbf{r}}.
\end{split}
\end{equation}
Using Eqs. (\ref{gyrotropic_force}-\ref{spin_current_force}) we can rewrite LLG equation (\ref{LLG_torque_current_thermal}) in the form of the Thiele equation
\begin{equation}\label{Thiel_torque_shot_2}
\begin{split}
&\widehat{G}\cdot\dot{\textbf{R}}+\alpha\widehat{D}\cdot\dot{\textbf{R}}+\textbf{F}_{eff}+\textbf{F}_\text{SOT}+\textbf{F}_{\delta h}+\textbf{F}_{\delta{I}}=0,
\end{split}
\end{equation}
where
\begin{equation}
\begin{split}
&G_{ij}=-\frac{1}{M_s^2\gamma}\int{\textbf{M}\cdot[\nabla_i\textbf{M}\times}\nabla_j\textbf{M}]d\textbf{r}, ~\frac{[\text{A m}^{-1}\cdot\text{m}^{-1}\cdot\text{A m}^{-1}\cdot\text{m}^{-1}\cdot\text{A m}^{-1}\cdot\text{m}^3]}{[\text{A}^2~\text{m}^{-2}\cdot\text{A s kg}^{-1}]}=[\text{kg s}^{-1}],\notag\\
&D_{ij}=\frac{1}{M_s\gamma}\int{\nabla_i\textbf{M}\cdot\nabla_j\textbf{M}d\textbf{r}}, ~\frac{[\text{m}^{-1}\cdot\text{A m}^{-1}\cdot\text{m}^{-1}\cdot\text{A m}^{-1}\cdot\text{m}^3]}{[\text{A}~\text{m}^{-1}\cdot\text{A s kg}^{-1}]}=[\text{kg s}^{-1}],\notag \\
&\textbf{F}_{eff}=\int{\nabla_i\textbf{M}\cdot\textbf{H}_\text{eff}d\textbf{r}}, ~[\text{m}^{-1}\cdot\text{A m}^{-1}\cdot\text{T}\cdot\text{m}^3]=[\text{m}^{-1}\cdot\text{A m}^{-1}\cdot\text{kg s}^{-2}\text{A}^{-1}\cdot\text{m}^3]=[\text{N}],\notag\\
&\textbf{F}_\text{SOT}=\frac{1}{M_s^2\mathcal{V}}\int{[\nabla_i\textbf{M}\times\textbf{M}]\cdot\textbf{I}_sd\textbf{r}}+\frac{\beta}{M_s\mathcal{V}}\int{\nabla_i\textbf{M}\cdot\textbf{I}_sd\textbf{r}},\\&\frac{[\text{m}^{-1}\cdot\text{A m}^{-1}\cdot\text{A m}^{-1}\cdot\text{J}\cdot\text{m}^3]}{[\text{A}^2\text{m}^{-2}\cdot\text{m}^3]}+\frac{[\text{m}^{-1}\cdot\text{A m}^{-1}\cdot\text{J}\cdot\text{m}^3]}{[\text{A}\text{m}^{-1}\cdot\text{m}^3]}=[\text{N}],\notag\\ 
&\textbf{F}_{\delta h}=\int{\nabla_i\textbf{M}\cdot\delta\textbf{h}d\textbf{r}}, ~[\text{m}^{-1}\cdot\text{A m}^{-1}\cdot\text{T}\cdot\text{m}^3]=[\text{m}^{-1}\cdot\text{A m}^{-1}\cdot\text{kg s}^{-2}\text{A}^{-1}\cdot\text{m}^3]=[\text{N}],\notag\\
&\textbf{F}_{\delta{I}}=\frac{1}{M_s^2\mathcal{V}}\int{[\nabla_i\textbf{M}\times\textbf{M}]\cdot\delta\textbf{I}d\textbf{r}}
+\frac{\beta}{M_s\mathcal{V}}\int{\nabla_i\textbf{M}\cdot\delta\textbf{I}d\textbf{r}},\notag
\\&\frac{[\text{m}^{-1}\cdot\text{A m}^{-1}\cdot\text{A m}^{-1}\cdot\text{J}\cdot\text{m}^3]}{[\text{A}^2\text{m}^{-2}\cdot\text{m}^3]}+\frac{[\text{m}^{-1}\cdot\text{A m}^{-1}\cdot\text{J}\cdot\text{m}^3]}{[\text{A}\text{m}^{-1}\cdot\text{m}^3]}=[\text{N}].\notag
\end{split}
\end{equation}
Here and below, the physical units of the corresponding quantities are indicated in square brackets. 
\end{itemize}

\section{II. Fluctuation forces. Diffusion coefficients of magnetic textures}\label{Sup_2}

The Thiele equations (\ref{Thiel_torque_shot_2}) contain two fluctuation forces of different nature. The force $\textbf{F}_{\delta h}$ is caused by thermal fluctuations $\delta\textbf{h}$, while the force $\textbf{F}_{\delta{I}}$ is caused by spin current fluctuations $\delta\textbf{I}$. The correlation dependences of the indicated forces are derived from the known relationships for fluctuation fields $\delta\textbf{h}$ and $\delta\textbf{I}$,
\begin{equation}
\begin{split}\label{current_thermal_noise_}
&\langle{\delta{h}}_{i}(t,\textbf{r}){\delta{h}}_{j}(t',\textbf{r}')\rangle=2\mathcal{D}_{T}\delta_{ij}\delta(\textbf{r}-\textbf{r}')\delta(t-t')=\frac{2{\alpha}k_BT}{\gamma{M}_s}\delta_{ij}\delta(\textbf{r}-\textbf{r}')\delta(t-t'), \frac{[\text{J K}^{-1}\cdot\text{K}\cdot\text{m}^{-3}\cdot{s}^{-1}]}{[\text{A s kg}^{-1}\cdot\text{A m}^{-1}]}=[\text{T}^2]\\ &\langle{\delta{I}}_{i}(t,\textbf{r}){\delta{I}}_{j}(t',\textbf{r}')\rangle=2\mathcal{D}_{\text{I}}\delta_{ij}\delta(\textbf{r}-\textbf{r}')\delta(t-t')=\frac{\alpha{M}_s\mathcal{V}^2}{\gamma}\frac{\delta\varepsilon}{E_F}(eU\coth{\frac{eU}{2k_BT}}-2k_BT)\delta_{ij}\delta(\textbf{r}-\textbf{r}')\delta(t-t'), [\text{J}^2]\\
&\langle{\delta{h}}_{i}(t,\textbf{r}){\delta{I}}_{j}(t',\textbf{r}')\rangle=0.
\end{split}
\end{equation}
Following approach of Refs. \cite{Kamppeter_PRB_1999,Thiaville_2018} we derive the autocorrelation function of the force $F_{\delta\text{h},i}$ as
\begin{equation}\label{cor_F_zeta}
\begin{split}
&\langle{F}_{\delta\text{h},i}F_{\delta\text{h},j}\rangle=\left\langle\int{\nabla_i{{M}_m}\delta{h}_md\textbf{r}\int\nabla_j{{M}_n}\delta{h}_nd\textbf{r'}}\right\rangle=\int\int{\nabla_i{{M}_{m}}\nabla_j{{M}_{n}}\left\langle{\delta h}_m{\delta h}_n\right\rangle{d}\textbf{r}{d}\textbf{r'}}\\
&=2\mathcal{D}_{T}\delta_{mn}\delta(t-t')\int{\nabla_i{{M}_m}\nabla_j{{M}_n}{d}\textbf{r}}=2{\alpha}k_BT{D}_{ij}\delta(t-t').
\end{split}
\end{equation}
The thermal and spin-current noises are of multiplicative nature. However, removing the magnetization vector gradient from the averaging brackets is justified by the assumption that this vector does not undergo strong perturbations relative to the noise-free case \cite{Kamppeter_PRB_1999,Thiaville_2018}.
The autocorrelation function for the force ${F}_{\delta\text{I},i}$ reads
\begin{equation}\label{cor_F_d}
\begin{split}
&\langle{F}_{\delta\text{I},i}F_{\delta\text{I},j}\rangle=\frac{\beta^2}{M_s^2\mathcal{V}^2}\left\langle\int{\nabla_i{{M}_m}\cdot\delta{I}_md\textbf{r}\int\nabla_j{{M}_n}\cdot\delta{I}_nd\textbf{r'}}\right\rangle+\frac{2\beta}{M_s^3\mathcal{V}^2}\left\langle\int{\nabla_i{{M}_m}\cdot\delta{I}_md\textbf{r}\int[\nabla_j\textbf{M}\times\textbf{M}]_n\delta{I}_nd\textbf{r'}}\right\rangle \\
&+\frac{1}{M_s^4\mathcal{V}^2}\left\langle\int[\nabla_i\textbf{M}\times\textbf{M}]_m\cdot\delta{I}_md\textbf{r}\int[\nabla_j\textbf{M}\times\textbf{M}]_n\delta{I}_nd\textbf{r'}\right\rangle=\frac{\beta^2}{M_s^2\mathcal{V}^2}\int\int\nabla_i{{M}_m}\cdot\nabla_j{{M}_n}\left\langle\delta{I}_{m}\delta{I}_{n}\right\rangle{d}\textbf{r}{d}\textbf{r'}\\
&+\frac{2\beta}{M_s^3\mathcal{V}^2}\int\int\bcancel{\nabla_i{{M}_m}[\nabla_j\textbf{M}\times\textbf{M}]_n}\left\langle\delta{I}_{m}\delta{I}_{n}\right\rangle{d}\textbf{r}{d}\textbf{r'}+\frac{1}{M_s^4\mathcal{V}^2}\int\int[\nabla_i\textbf{M}\times\textbf{M}]_m[\nabla_j\textbf{M}\times\textbf{M}]_n\left\langle\delta{I}_{m}\delta{I}_{n}\right\rangle{d}\textbf{r}{d}\textbf{r'} \\
&=\frac{2\mathcal{D}_{\text{I}}\delta_{mn}\delta(t-t')}{M_s^4\mathcal{V}^2}\int[\nabla_i\textbf{M}\times\textbf{M}]_m[\nabla_j\textbf{M}\times\textbf{M}]_n{d}\textbf{r}+\frac{2\mathcal{D}_{\text{I}}\beta^2\delta_{mn}\delta(t-t')}{M_s^2\mathcal{V}^2}\int\nabla_i{{M}_m}\cdot\nabla_j{{M}_n}{d}\textbf{r}\\
&=\frac{2\mathcal{D}_{\text{I}}\delta(t-t')}{M_s^4\mathcal{V}^2}\int\nabla_i\textbf{M}\cdot(\nabla_j\textbf{M}{M}^2_s-\textbf{M}(\bcancel{\nabla_j\textbf{M}\cdot\textbf{M}})){d}\textbf{r}+\frac{2\mathcal{D}_{\text{I}}\beta^2\delta(t-t')}{M_s^2\mathcal{V}^2}\int\nabla_i{\textbf{M}}\cdot\nabla_j{\textbf{M}}{d}\textbf{r}\\
&=\frac{2\mathcal{D}_{\text{I}}(1+\beta^2)\delta(t-t')}{M_s^2\mathcal{V}^2}\int\nabla_i{\textbf{M}}\cdot\nabla_j{\textbf{M}}{d}\textbf{r}=\alpha(1+\beta^2)\frac{\delta\varepsilon}{E_F}(eU\coth{\frac{eU}{2k_BT}}-2k_BT)D_{ij}\delta(t-t').
\end{split}
\end{equation}
It is easy to see that cross-correlations with thermal and spin current fluctuation forces are zero, $\langle{F}_{\delta\text{h},i}F_{\delta\text{I}_\text{s},j}\rangle=0$. The correlation functions for equilibrium and nonequilibrium stochastic forces are

\begin{equation}\label{cor_forces}
\begin{split}
&\langle{F}_{\delta\text{h},i}F_{\delta\text{h},j}\rangle=2{\alpha}k_BT{D}_{ij}\delta(t-t'), \\
&\langle{F}_{\delta\text{I},i}F_{\delta\text{I},j}\rangle=\alpha(1+\beta^2)\frac{\delta\varepsilon}{E_F}(eU\coth{\frac{eU}{2k_BT}}-2k_BT)D_{ij}\delta(t-t'),\\
&\langle{F}_{\delta\text{I},i}F_{\delta\text{h},j}\rangle=0
\end{split}
\end{equation}
In the case of two-dimensional magnetic films, we can write Eq. (\ref{Thiel_torque_shot_2}) as
\begin{equation}\label{diff_equation}
\begin{split}
&-G_{xy}\dot{R}_y+\alpha{D}_{xx}\dot{R}_x+\alpha{D}_{xy}\dot{R}_y+F_x=0, \\
&G_{xy}\dot{R}_x+\alpha{D}_{xy}\dot{R}_x+\alpha{D}_{yy}\dot{R}_y+F_y=0,
\end{split}
\end{equation}
Here $\textbf{F}=\textbf{F}_{eff}+\textbf{F}_\text{SOT}+\textbf{F}_{\delta h}+\textbf{F}_{\delta{I}}$. Next, solving linear algebraic equations (\ref{diff_equation}), we find velocity $\textbf{v}(t)=\{v_x(t),v_y(t)\}=\dot{\textbf{R}}(t)$,
\begin{equation}\label{velocity}
\begin{split}
&\dot{R}_x=v_x=\frac{-G_{xy}F_y + \alpha{D}_{xy}F_y - \alpha{D}_{yy}F_x}{-\alpha^2D_{xy}^2 + G_{xy}^2 + \alpha^2D_{xx}D_{yy}}, \\
&\dot{R}_y=v_y=\frac{G_{xy}F_x+\alpha{D}_{xy}F_x-\alpha{D}_{xx}F_y}{-\alpha^2D_{xy}^2 + G_{xy}^2 + \alpha^2D_{xx}D_{yy}}.
\end{split}
\end{equation}
\subsection{Skirmion diffusion constant}
The resulting Eqs. (\ref{cor_forces}) and (\ref{diff_equation}) make it possible to describe the diffusion of magnetic textures in thin ferromagnetic films. Because of the revolution symmetry of a skyrmion at rest, we can assume that ${D}_{xy}=0$, ${D}_{xx}={D}_{yy}=D$ and ${G}_{xy}=G$ \cite{Thiaville_2018}, which is also confirmed by our micromagnetic simulation. Next, we calculate the velocity $\textbf{v}(t)$ of the skyrmion as
\begin{equation}\label{velocity_skyrmion}
\begin{split}
&v_x=\frac{-GF_y - \alpha{D}F_x}{G^2 + \alpha^2D^2}, \\
&v_y=\frac{GF_x-\alpha{D}F_y}{G^2 + \alpha^2D^2}.
\end{split}
\end{equation}
The correlation functions for the components of the skyrmion velocity are

\begin{equation}\label{cor_skyrmion}
\begin{split}
&\langle{v}_x(t){v}_x(t')\rangle=\frac{\langle-{G}F_y(t) - \alpha{D}F_x(t),-{G}F_y(t') - \alpha{D}F_x(t')\rangle}{(G^2 + \alpha^2D^2)^2}=\frac{{G}^2\langle{F}_y(t){F}_y(t')\rangle + \alpha^2{D}^2\langle{F}_x(t){F}_x(t')\rangle}{(G^2 + \alpha^2D^2)^2} \\
&=\frac{{G}^2(\langle{F}_{h,y}(t){F}_{h,y}(t')\rangle+\langle{F}_{\text{I},y}(t){F}_{\text{I},y}(t')\rangle) + \alpha^2{D}^2(\langle{F}_{h,x}(t){F}_{h,x}(t')\rangle+\langle{F}_{\text{I},x}(t){F}_{\text{I},x}(t')\rangle)}{(G^2 + \alpha^2D^2)^2} \\
&=\frac{({G}^2+\alpha^2{D}^2)(2k_BT+(1+\beta^2)\frac{\delta\varepsilon}{E_F}(eV\coth{\frac{eV}{2k_BT}}-2k_BT))\alpha{D}\delta(t-t')}{(G^2 + \alpha^2D^2)^2} \\ 
&=\frac{\alpha{D}(2k_BT+(1+\beta^2)\frac{\delta\varepsilon}{E_F}(eV\coth{\frac{eV}{2k_BT}}-2k_BT))\delta(t-t')}{G^2 + \alpha^2D^2},\\
&\langle{v}_y(t){v}_y(t')\rangle=\frac{\langle{G}F_x(t) - \alpha{D}F_y(t),{G}F_x(t') - \alpha{D}F_y(t')\rangle}{(G^2 + \alpha^2D^2)^2}=\frac{{G}^2\langle{F}_x(t){F}_x(t')\rangle + \alpha^2{D}^2\langle{F}_y(t){F}_y(t')\rangle}{(G^2 + \alpha^2D^2)^2} \\
&=\frac{{G}^2(\langle{F}_{h,x}(t){F}_{h,x}(t')\rangle+\langle{F}_{\text{I},x}(t){F}_{\text{I},x}(t')\rangle) + \alpha^2{D}^2(\langle{F}_{h,y}(t){F}_{h,y}(t')\rangle+\langle{F}_{\text{I},y}(t){F}_{\text{I},y}(t')\rangle)}{(G^2 + \alpha^2D^2)^2} \\
&=\frac{({G}^2+\alpha^2{D}^2)(2k_BT+(1+\beta^2)\frac{\delta\varepsilon}{E_F}(eV\coth{\frac{eV}{2k_BT}}-2k_BT))\alpha{D}\delta(t-t')}{(G^2 + \alpha^2D^2)^2} \\ 
&=\frac{\alpha{D}(2k_BT+(1+\beta^2)\frac{\delta\varepsilon}{E_F}(eV\coth{\frac{eV}{2k_BT}}-2k_BT))\delta(t-t')}{G^2 + \alpha^2D^2}, \\
&\langle{v}_x(t){v}_y(t')\rangle=\frac{\langle-{G}F_y(t) - \alpha{D}F_x(t),{G}F_x(t') - \alpha{D}F_y(t')\rangle}{(G^2 + \alpha^2D^2)^2}=\frac{\alpha{G}D\langle{F}_y(t){F}_y(t')\rangle - \alpha{G}{D}\langle{F}_x(t){F}_x(t')\rangle}{(G^2 + \alpha^2D^2)^2} \\
&=\frac{-\alpha{G}D(\langle{F}_{h,x}(t){F}_{h,x}(t')\rangle+\langle{F}_{\text{I},x}(t){F}_{\text{I},x}(t')\rangle) + \alpha{G}{D}(\langle{F}_{h,y}(t){F}_{h,y}(t')\rangle+\langle{F}_{\text{I},y}(t){F}_{\text{I},y}(t')\rangle)}{(G^2 + \alpha^2D^2)^2}=0.
\end{split}
\end{equation}
The time integration $R_x=\int_0^t{v_x(t')dt'}$ yields the mean squared displacements
\begin{equation}\label{skyrmion_diffusion}
\begin{split}
&\langle{R}^2_x(t)\rangle=\langle{R}^2_y(t)\rangle=2\mathcal{D}t=\frac{\alpha{D}\left(2k_BT+(1+\beta^2)\frac{\delta\varepsilon}{E_F}(eV\coth{\frac{eV}{2k_BT}}-2k_BT)\right)}{G^2 + \alpha^2D^2}t.
\end{split}
\end{equation}
The diffusion relations for skyrmion (\ref{skyrmion_diffusion}) correspond to the known results  \cite{Schutte_PRB_2014,Thiaville_2018} (thermal noise only) in the absence of an external current source at $V\rightarrow0$.

\section{III. Micromagnetic simulation}\label{Sup_3}

\begin{figure}
\centering
\includegraphics[width=0.7\linewidth]{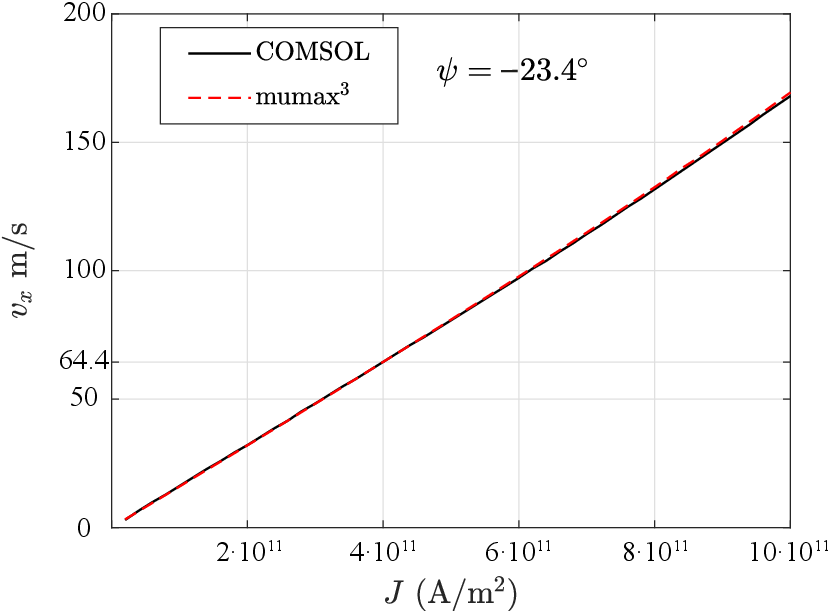}
\caption{\label{vel_skyrm} Velocity of skyrmion $v_x$ versus current density $J$ obtained in COMSOL Multiphysics (black solid line) and mumax$^3$ (red dotted line).}
\end{figure}
Micromagnetic simulations were performed with COMSOL Multiphysics using the micromagnetic module \cite{comsolMicromag,micromagModule} for two-dimensional system. This multiphysics platform allows for the generalization of the LLG equations by incorporating additional nonequilibrium fluctuation terms. The ability of COMSOL Multiphysics platform to extend the functionality of the standard micromagnetic model is particularly relevant for studying complex systems. 
In order to compare the results obtained using micromagnetic modeling in COMSOL with those obtained using the commonly used open-source platform mumax$^3$, we establish the relationship between the corresponding coefficients.
Namely, the effective field in mumax$^3$ has the form
\begin{equation}
\begin{split}\label{effective_field_mumax}
\textbf{H}_\text{eff}=\frac{2A_\text{ex}}{M_s}\nabla^2\textbf{m}+\frac{2K_u}{M_s}m_z\textbf{e}_z+\frac{2D_\text{DMI}}{M_s}(\nabla{m}_z-(\nabla\cdot\textbf{m})\textbf{e}_z),~[T],
\end{split}
\end{equation}

while the effective field in COMSOL Multiphysics is
\begin{equation}
\begin{split}\label{effective_field_COMSOL}
\frac{\textbf{H}_\text{eff}}{\mu_0}={A'_\text{ex}}\nabla^2\textbf{m}+{K'_u}m_z\textbf{e}_z-{D'_\text{DMI}}(\nabla{m}_z-(\nabla\cdot\textbf{m})\textbf{e}_z),~[\text{Am}^{-1}].
\end{split}
\end{equation}
In the main text of the paper we use the coefficients adopted for mumax$^3$. Consequently, the relationship between the coefficients of the two numerical platforms is
\begin{equation}
A'_\text{ex}=\frac{2A_\text{ex}}{M_s\mu_0}, \quad K'_\text{u}=\frac{2K_\text{ex}}{M_s\mu_0}, \quad D'_\text{DMI}=\frac{2D_\text{DMI}}{M_s\mu_0}.
\end{equation}
We used the following parameters in our simulations: The exchange coefficient $A_\text{ex}=10^{-11}$~J/m, the anisotropy $K_u=10^6$~J/m$^3$, the saturation magnetization $M_s=0.6\cdot10^{6}$~A/m, the interfacial DMI coefficient $D_\text{DMI}=-3.5\cdot10^{-3}$~J/m$^2$ and the damping parameters $\alpha=\beta=0.3$. The racetrack sample has ($x$-$y$) plane section of $500\times150$~nm$^2$ and thickness $\Delta_f=1$ nm. Electric current is $J=4\cdot10^{11}$~A/m$^2$,  applied voltage is $U=0.5$~V and temperature is $T=300$~K.  Such parameters are typical for the experiments on the racetracks \cite{Sampaio2013,JaeChun_ACSNANO_2024,JaeChun_Science_2024}. The topological properties of skyrmions can be completely characterized by three different numbers ($Q, Q_v, Q_h$)  denoting the topological charge, vorticity number and helicity number, respectively \cite{Tretiakov_PhysRevB_2017,Zhang_2020}. We restrict ourselves to the DMI, which leads to stable N\'eel-type skyrmions ($1,1,\pi$). 

Next, we compare both approaches to micromagnetic modeling, namely COMSOL Multiphysics and mumax$^3$ platforms, for the parameters presented above. As shown in Fig. \ref{vel_skyrm}, good agreement was obtained between the skyrmion velocity $v_x$ on the racetrack as a  function of the current density $J$. It is also worth to note that the simulation were performed for skyrmion dynamics with suppression of the skyrmion Hall effect. Such suppression is obtained by choosing a polarization angle as $\psi=-23.4^{\circ}$. The results obtained indicate the equivalence of the micromagnetic modeling based on COMSOL Multiphysics and mumax$^3$ platforms.

Once we have determined all the modeling parameters, we can evaluate the parameters G and D in Eq. (\ref{velocity_skyrmion}) as 
\begin{equation}
\begin{split}
&G=\frac{M_s\Delta_f}{\gamma}\int{\textbf{m}\cdot[\nabla_x\textbf{m}\times}\nabla_y\textbf{m}]dxdy=4\pi{M}_s\Delta_f/\gamma= 4.29\cdot10^{-14}~[kg/s], \\
&D=\frac{M_s\Delta_f}{\gamma}\int{\nabla_x\textbf{m}\cdot\nabla_x\textbf{m}dxdy}=18.1M_s\Delta_f/\gamma= 6.17\cdot10^{-14}~[kg/s],
\end{split}
\end{equation}
and the mean velocities along the $x$ and  $y$ axes as
\begin{equation}
\begin{split}
&\langle{v}_x\rangle\approx\frac{-GF_{\text{SOT},y} - \alpha{D}F_{\text{SOT},x}}{G^2 + \alpha^2D^2}=\frac{Pj\hbar}{2e}\cdot\frac{-G\int{[\nabla_y\textbf{m}\times\textbf{m}]\cdot\textbf{p}dxdy} - \alpha{D}\int{[\nabla_x\textbf{m}\times\textbf{m}]\cdot\textbf{p}dxdy}}{G^2 + \alpha^2D^2}=64.4 ~[m/s],\\
&\langle{v}_y\rangle\approx\frac{GF_{\text{SOT},x}-\alpha{D}F_{\text{SOT},y}}{G^2 + \alpha^2D^2}=\frac{Pj\hbar}{2e}\cdot\frac{G\int{[\nabla_x\textbf{m}\times\textbf{m}]\cdot\textbf{p}dxdy}-\alpha{D}\int{[\nabla_y\textbf{m}\times\textbf{m}]\cdot\textbf{p}dxdy}}{G^2 + \alpha^2D^2}=0.
\end{split}
\end{equation}
Here $\langle{v}_x\rangle$ and $\langle{v}_y\rangle$ depend mainly on the force induced by Slonczewski's torque term $F_{\text{SOT},i}=\frac{Pj\hbar}{2e}\int{[\nabla_i\textbf{m}\times\textbf{m}]\cdot\textbf{p}dxdy}$. It should also be noted that the vector $\textbf{m}$ in the integrand represents the magnetization field distribution for an equilibrium isolated skyrmion on a given racetrack.

\section{IV. First Passage Time moments of stochastic equations}\label{Sup_4}

Application of the Thiele approach to the LLG equation with noise terms and suppressed skyrmion Hall effect leads to a system of two independent stochastic equations with constant coefficients:
\begin{equation}\label{equSDE2D}
\begin{split}
&\dot{\textbf{R}}=\langle\textbf{v}\rangle+\sqrt{2\mathcal{D}}\bm{\xi}(t).
\end{split}
\end{equation}
Here,  $\langle\textbf{v}\rangle$ is the mean drift velocity, whose components are determined from Eq.~(\ref{velocity_skyrmion}), and $\bm{\xi}(t)$ is a white Gaussian noise with the zero mean and autocorrelation function $\langle\xi_i(t)\xi_j(t)\rangle=\delta_{ij}\delta(t-t')$.  This stochastic equation describes the classical diffusion process with drift and has a well-known analytical solution \cite{CoxMiller_2001}.  

We treat the FPT problem componentwise. For each component, the moments of the FPT $\tau$ to an absorbing boundary satisfy the standard hierarchy of ODEs derived from the backward Kolmogorov equation (see §5.10 in \cite{CoxMiller_2001}):
\begin{equation}\label{equMomentsOfFPTequ}
\mathcal{D} \frac{\mathrm{d}^2}{\mathrm{d}\lambda^2}\langle \tau_{n}(\lambda)\rangle + \langle v_\lambda \rangle\frac{\mathrm{d}}{\mathrm{d}\lambda}\langle \tau_{n}(\lambda)\rangle=-n\langle \tau_{n-1}(\lambda)\rangle
\end{equation}
Here $\lambda = \{x,y\}$,  $\langle \tau_{n}(\lambda)\rangle$ is $n$-th moment of  the first passage time $\tau$ and $\langle v_\lambda \rangle$ is the drift velocity along $\lambda$-axis. The variable $\lambda$ belongs to the range  $\lambda_a \leq\lambda \leq \lambda_b $ with absorbing boundaries at the points $\lambda_a$ and $\lambda_b$, which correspond to the conditions $\langle \tau_{n}(y)\rangle|_{y=\lambda_a} = \langle \tau_{n}(y)\rangle|_{y=\lambda_b} = 0$.  Using zero-moment condition $\langle \tau_{0}(\lambda)\rangle = 1$, one has the following system of equations for the first two moments:
\begin{equation}\label{equTwoMomentsOfFPTequ}
\begin{cases}
\mathcal{D} \frac{\mathrm{d}^2}{\mathrm{d}\lambda^2}\langle \tau_{1}(\lambda)\rangle + \langle v_\lambda \rangle\frac{\mathrm{d}}{\mathrm{d}\lambda}\langle \tau_{1}(\lambda)\rangle=-1 \\
\mathcal{D} \frac{\mathrm{d}^2}{\mathrm{d}\lambda^2}\langle \tau_{2}(\lambda)\rangle + \langle v_\lambda \rangle\frac{\mathrm{d}}{\mathrm{d}\lambda}\langle \tau_{2}(\lambda)\rangle=-2\langle \tau_{1}(\lambda)\rangle 
\end{cases}
\end{equation}
which can be easily solved with specified boundary conditions for constant  $\mathcal{D}$ and $\langle v_\lambda \rangle$.

We set the $x$-axis along the racetrack with the width 2$a$. Skyrmions start at the origin of the coordinate system $x=y=0$, and move with a drift velocity $\langle v_x \rangle$ along the racetrack. 

For the problem of first passage time to the side racetrack boundary, we have $\langle v_y \rangle = 0$,  and  the boundary conditions $\langle \tau_{1,2}(y)\rangle|_{y=-a} = \langle \tau_{1,2}(y)|_{y=a}\rangle = 0$.  In this case, the solution of  the system  \eqref{equTwoMomentsOfFPTequ} leads to the following expressions for the mean $\langle \tau_y\rangle$ and the standard deviation $\sigma_y$ of  FPT to the side boundary:
\begin{equation}\label{equFPTacross}
\begin{split}
\langle \tau_y\rangle &=\frac{a^2}{2\mathcal{D}} \\
\sigma_y  &=  \sqrt{\langle\tau^2_{y}\rangle - \langle\tau_{y}\rangle^2}=\frac{a^2}{\sqrt{6}\mathcal{D}}
\end{split}
\end{equation}

For the case of the FPT along the racetrack to the detection area at a distance $d$, the boundary conditions are $\langle \tau_{1,2}(x)\rangle|_{x=-\infty} = \langle \tau_{1,2}(x)|_{x=d}\rangle = 0$. In that case, the mean $\langle \tau_x\rangle$ and the standard deviation $\sigma_x$ of  FPT from the starting position to the detection area on the racetrack read
\begin{equation}\label{equFPTalong}
\begin{split}
\langle \tau_x\rangle  &=\frac{d}{\left<v_x\right>}, \\
  \sigma_x  &= \sqrt{\langle\tau^2_{x}\rangle - \langle\tau_{x}\rangle^2}=\sqrt{\frac{2\mathcal{D}d}{\left<v_x\right>^3}}.
\end{split}
\end{equation}

\end{document}